\documentclass[aps,twocolumn, superscriptaddress]{revtex4-1}
\usepackage[pdftex]{graphicx}
 \usepackage{amsmath}
\usepackage[T1]{fontenc}
\usepackage{amsfonts}
\usepackage{lipsum}
\usepackage{amssymb, mathrsfs}
\usepackage{braket}
\usepackage{subfigure}
\def\beq{\begin{equation}}
\def\eeq{\end{equation}}
\def\bsp{\begin{split}}
\def\esp{\end{split}}
\def\bea{\begin{eqnarray}}
\def\eea{\end{eqnarray}}
\def\ba{\begin{array}}
\def\ea{\end{array}}

\def\lb{\left(}
\def\rb{\right)}

\def\l.{\left.}
\def\r.{\right.}

\def\ra{\rangle}
\def\la{\langle}

\begin{document}

\title{Topological Magnetic Excitations  on the Distorted Kagom\'e Antiferromagnets:\\ Applications to  Volborthite, Vesignieite, and Edwardsite.}
\author{S. A. Owerre}
\affiliation{Perimeter Institute for Theoretical Physics, 31 Caroline St. N., Waterloo, Ontario N2L 2Y5, Canada.}
\affiliation{African Institute for Mathematical Sciences, 6 Melrose Road, Muizenberg, Cape Town 7945, South Africa.}

\begin{abstract}
  In this Letter, we identify a noncoplanar chiral spin texture on the distorted  kagom\'e-lattice antiferromagnets  induced by   the presence of a magnetic field applied perpendicular to the distorted kagom\'e plane. The noncoplanar chiral spin texture has a nonzero scalar spin chirality similar to a chiral quantum spin liquid with broken  time-reversal symmetry. In the noncoplanar regime we observe nontrivial  topological magnetic excitations and thermal Hall response through the emergent field-induced scalar spin chirality in contrast to collinear ferromagnets in which the   Dzyaloshinskii-Moriya (DM) spin-orbit interaction is the driving force.   Our results shed light on recent observation of  thermal Hall conductivity in distorted  kagom\'e volborthite at nonzero magnetic field with no signal of DM spin-orbit interaction, and the results also apply to other distorted kagom\'e antiferromagnets such as vesignieite and edwardsite.
 \end{abstract}
\maketitle

\section{Introduction}
A chiral spin liquid (CSL)  is a particular type of spin liquids in which time-reversal symmetry (TRS) is broken macroscopically  \cite{kal,bas,wen}. The scalar spin chirality $\la\chi_{ijk}\ra= \la{\bold S}_i\cdot\lb \bold S_{j}\times\bold S_{k}\rb\ra$  measures broken TRS and defines  the hallmark of a CSL.    It is also generally believed that the magnetic excitations in any quantum spin liquid (QSL) are  bosonic spin-$1/2$ spinons in which the conventional bosonic spin-$1$ magnons fractionalize. According to Kalmeyer and Laughlin \cite{kal} a CSL can be thought of as a bosonic version of $\nu=1/2$ Laughlin state in fractional quantum Hall effect.  In recent years, a plethora of theoretical models has been proposed in which a CSL phase can be stabilized. The recent theoretical numerical work shows that a CSL can emerge in a kagom\'e lattice Mott insulator in which a magnetic field induces an explicit spin chirality interaction in a $t/U$ perturbative expansion of the  Hubbard model at half filling \cite{bau}, hence TRS is broken explicitly. In other numerical works a spontaneously broken TRS  CSL is found away from the isotropic kagom\'e-lattice antiferromagnet by adding additional second and third nearest neighbour interactions \cite{shou,shou1,shou2,shou3, mes2} or Ising anisotropy \cite{shou4}.

In real materials, however, a CSL has remained elusive because material  synthesis usually comes with defects such as structural distortion and perturbative anisotropy like the intrinsic DM spin-orbit interaction \cite{dm,dm1}, which tend to destabilize  disordered QSL and induce a magnetic long-range order. For quantum kagom\'e antiferomagnets (QKAF), various experimentally accessible materials such as iron jarosites \cite{sup1a,men4a, dan, men1}, vesignieite BaCu$_3$V$_2$O$_8$(OH)$_2$ \cite{oka,oka1,oka2,okaa,okaa1, men0, men2},  edwardsite  Cd$_2$Cu$_3$(SO$_4$)$_2$(OH)$_6$.$4$H$_2$O  \cite{oka3}, and volborthite  Cu$_3$V$_2$O$_7$(OH)$_2$$\cdot$2H$_2$O \cite{wat1,Yo,Yo1,Yo2} have long-range magnetic orders below certain temperatures and  they are generally attributed  to the presence of DM spin-orbit interaction. The recent proposals of  QSL materials are  herbertsmithite ZnCu$_3$(OH)$_6$Cl$_2$ \cite{nor,zor} and the calcium-chromium oxide  Ca$_{10}$Cr$_7$O$_{28}$ \cite{balz}  which show a continuum of spinon excitations in inelastic neutron scattering experiment \cite{tia, balz}. Nevertheless, an applied magnetic field or pressure destabilizes the  QSL  nature of these materials and leads to  an induced magnetic order at low temperatures \cite{jeo,koz,balz}.

The emerging experimental technique to probe the nature of magnetic excitations in quantum magnets with possible QSL  ground states is the measurement of thermal Hall effect. For quantum magnets with magnetic long-range order the thermal Hall effect has been realized  in the collinear  kagom\'e ferromagnet Cu(1-3, bdc) \cite{alex6}  and  a number of collinear  pyrochlore ferromagnets \cite{alex1,alex1a}. In this case the transverse thermal Hall conductivity ($\kappa_{xy}$) can be explained in terms of Berry curvature induced by the DM spin-orbit interaction \cite{alex0,alex2, shin1} leading to topological magnons  \cite{alex4,alex44, zhh,alex6a,sol,sol1,kwon, jun,chen} and Weyl magnons \cite{su,mok} similar to  spin-orbit coupling electronic systems \cite{kane,kane1, aab1,aab2,aab3}. In a recent experiment \cite{wat}, a nonzero $\kappa_{xy}$  has been observed  in a frustrated distorted kagom\'e volborthite at a strong magnetic field of  $15 ~\text{T}$  with no signs of the DM spin-orbit interaction and no discernible thermal Hall signal was observed at zero magnetic field \cite{fot}. The authors attributed the presence of $\kappa_{xy}$ to nontrivial elementary excitations in the gapless QSL phase, however a strong field of $15 ~\text{T}$  causes  low-temperature magnetic phases in volborthite  \cite{Yo,Yo1, Yo2}. Nevertheless, the exact nature of the low-temperature magnetic phases at  $15 ~\text{T}$ is poorly understood, but the intrinsic DM spin-orbit anisotropy suggests a $\bold {Q=0}$ coplanar/noncollinear N\'eel order.

In previous exact  diagonalization study, it has been established that the out-of-plane DM spin-orbit anisotropy can induce  a quantum critical point (QCP)  in  spin-$1/2$ QKAF  at  $D_{\perp}^c/J\sim 0.1$  \cite{men3}. A moment free QSL phase is predicted for  $D_{\perp}<D_{\perp}^c$ and  a $\bold {Q=0}$ N\'eel phase exists for $D_{\perp}>D_{\perp}^c$.  From different experimental inspections, it is generally believed that both edwardsite and vesignieite are located in the ordered regime below certain temperature  with $0.1<D_\perp/J<0.16$ \cite{oka1,okaa1},  and there is a possibility that volborthite  also belongs to the ordered regime at low temperatures. But herbertsmithite ZnCu$_3$(OH)$_6$Cl$_2$ has $D_{\perp}/J\sim 0.08$ \cite{zor}, thus  remain a QSL. An applied  magnetic field (H $\perp$ 2D plane) can induce noncoplanar spin textures with a nonzero scalar spin chirality. The emergent scalar spin chirality due to noncoplanar spins might be different from the spontaneous scalar spin chirality in the CSL, but their effects  on the magnetic excitations should be the same. In this Letter we show that a nonzero Berry curvature can be induced by a chiral  spin texture in real space rather than the DM spin-orbit interaction.  The Berry curvature stems from the emergent scalar spin chirality  which measures the solid angle subtended by three noncoplanar spins in a unit triangular plaquette of the kagom\'e lattice. The Berry curvature of the chiral spin texture is the driving force of the thermal Hall effect in frustrated distorted QKAF with/without magnetic long-range order.

\section{Model}

 We consider the microscopic spin Hamiltonian on the distorted QKAF subject to a magnetic field perpendicular to the kagom\'e plane. The Hamiltonian is given by

\begin{figure}
\centering
\includegraphics[width=3in]{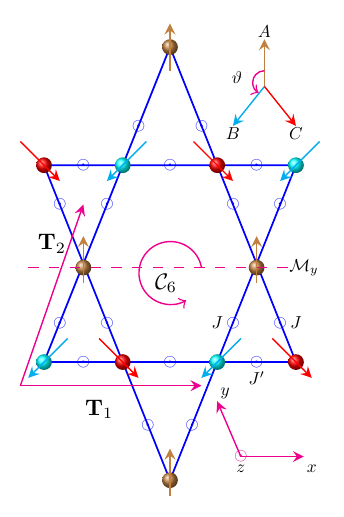}
\caption{Color online. The canted noncollinear/coplanar magnetic order with positive vector chirality on the distorted kagom\'e lattice and the symmetry group of the lattice. The out-of-plane DM interaction lies at the midpoints between two magnetic ions as indicated by small dotted circles. The spin triad are separated by an angle  $\vartheta\neq 120^\circ$ on each isosceles triangle with $J\neq J^\prime$ as shown at the top figure. }
\label{dis}
\end{figure}

 \begin{align}
\mathcal H&= \sum_{ \la ij\ra}\big [ J_{ij}{\bf S}_{i}\cdot{\bf S}_{j}+ { \bf D}_{ij}\cdot{ \bf S}_{i}\times{\bf S}_{j}\big]-{\bf H}\cdot\sum_{i} {\bf S}_{i},
\label{model}
\end{align}
where ${\bf S}_{i}$ is the magnetic spin moment at site $i$, and the summation runs over  nearest-neighbour spins.  $J_{ij}=J>0$ on the diagonal bonds, and $J_{ij}=J^\prime=J\delta$ on the horizontal bonds with $\delta \neq 1$  as shown in Fig.~\ref{dis}.   The inversion symmetry breaking between the bonds on the kagom\'e lattice allows  a DM vector ${ \bf D}_{ij}=-D_\perp\bold {\hat z}$  and they lie at the midpoints between two magnetic ions as shown in Fig.~\ref{dis}.  The last term is the out-of-plane external  magnetic field $\bold H=h\bold{\hat z}$ and $h=g\mu_B H$. The Hamiltonian \eqref{model}  is applicable to volborthite \cite{Yo4,fa, lii, apel,par} and other forms of Hamiltonian have also been proposed for volborthite \cite{Yo3a,andr,Yo3}.  The out-of-plane DM interaction is always present on the kagom\'e lattice and it stabilizes the coplanar/noncollinear spin configuration \cite{men1}.  In some kagom\'e materials  an in-plane DM interaction $D_{\parallel}$ may be present due to lack of mirror planes. It leads to  weak out-of-plane ferromagnetism with small ferromagnetic moment. However, it is usually negligible compared to the out-of-plane component for most spin-$1/2$ kagom\'e antiferromagnetic materials \cite{men3,zor} and can be removed by gauge transformation. For kagom\'e volborthite \cite{wat}, there was no signal of both in-plane and out-of-plane DM interaction on the observed $\kappa_{xy}$. Also, for potassium Fe-jarosite with spin-$5/2$ the in-plane DM interaction (or the DM interaction in general) does not necessarily induce topological magnetic excitations \cite{ men4a}.  Therefore,  we  will neglect the small in-plane DM component  at the moment and comment on its effects in the subsequent sections.


\begin{figure}
\centering
\includegraphics[width=3in]{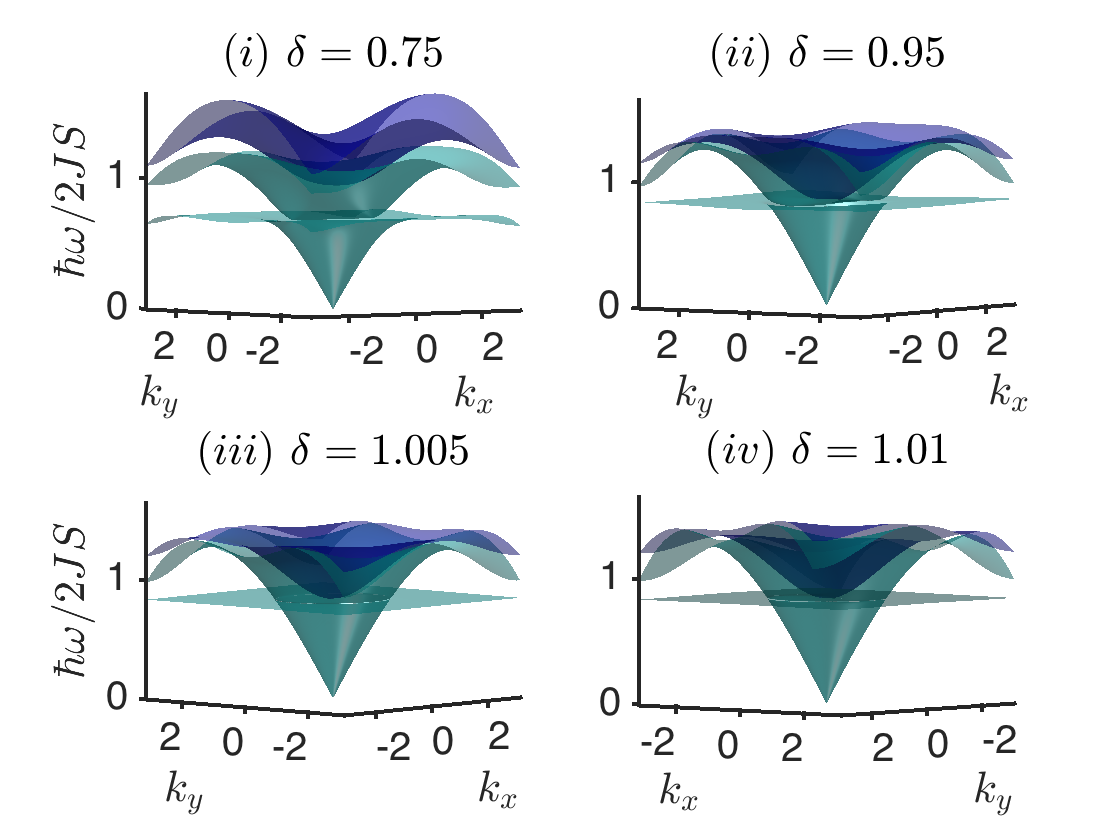}
\caption{Color online.  Magnon dispersion of distorted QKAF at zero magnetic field  and several values of distortion with $D_\perp/J=0.2$.}
\label{band}
\end{figure}

\begin{figure}
\centering
\includegraphics[width=3in]{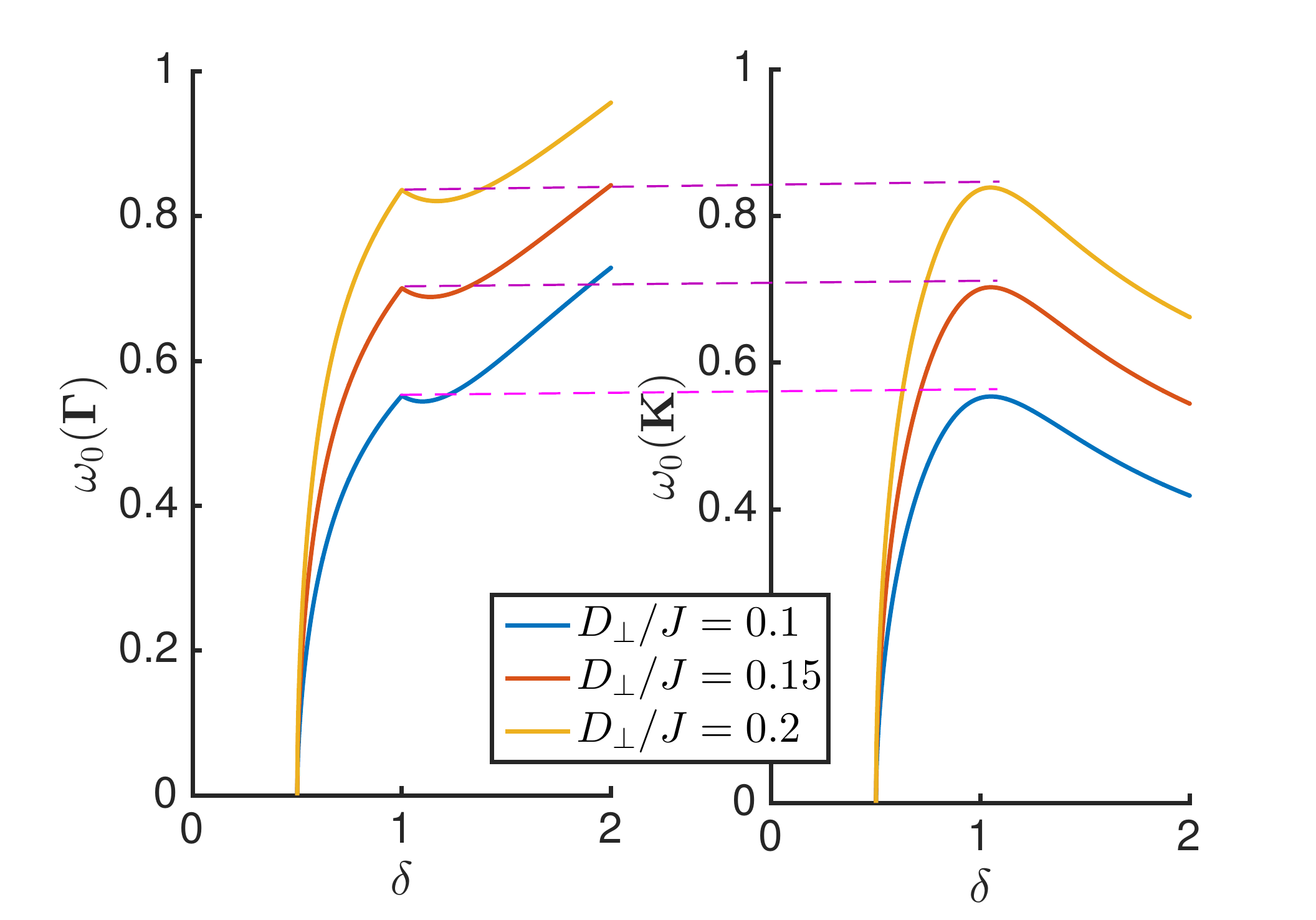}
\caption{Color online.   The lifted zero mode $\omega_0$ varies in the Brillouin zone as function of the distortion  for zero magnetic field $\mathcal H_{\phi}=0$. The dash lines connect isotropic point $\delta=1$. The coplanar/noncollinear order is valid for $\delta>0.5$.}
\label{E0}
\end{figure}

\begin{figure}
\centering
\includegraphics[width=3in]{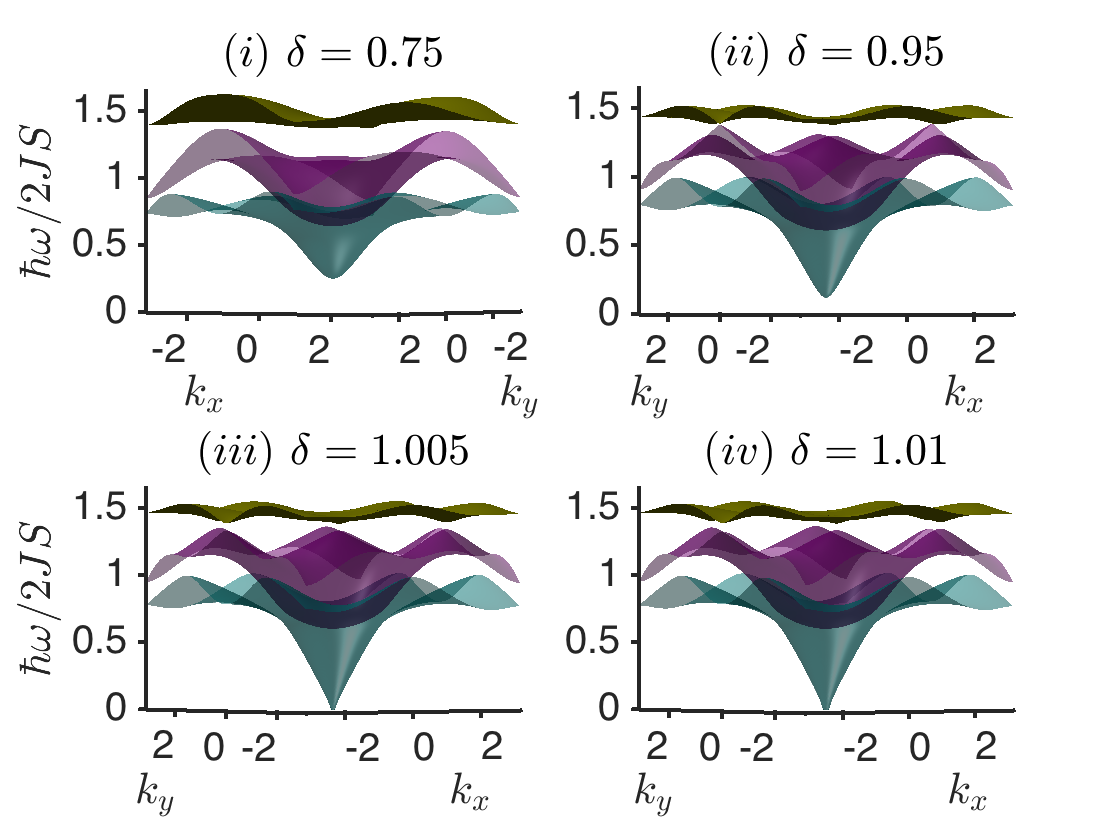}
\caption{Color online. Topological magnon dispersion of distorted QKAF at finite magnetic-field-induced  scalar spin chirality. The parameters chosen are $h/h_s=0.3$ and $D_\perp/J=0.2$.}
\label{band1}
\end{figure}

Let us start from what is known for distorted kagom\'e antiferromagnets at   ${\bf  D}_{ij}=h=0$ \cite{fa}. In this case, the Hamiltonian can be written as
\begin{align}
\mathcal H=\frac{J\delta}{2}\sum_{\Delta} \lb \frac{1}{\delta}{\bf S}_A+{\bf  S}_B+{\bf  S}_C\rb^2-\text{const.},
\end{align}
where ${ \bf S}_{A,B,C}$ form a triad as depicted in Fig.~\ref{dis}. It is easily seen that the classical energy is minimized for $\frac{1}{\delta}{\bf  S}_A+{\bf  S}_B+{\bf  S}_C=0$, yielding $\vartheta= \arccos(-1/2\delta)\neq 120^\circ$ for $\delta \neq 1$. The classical configuration is now a canted coplanar/noncollinear spins for $\delta >1/2$  \cite{fa}.  However, it also has an extensive degeneracy as in the ideal kagom\'e Heisenberg antiferromagnets. There are several limiting cases in which the system is either bipartite with collinear magnetic order or non-bipartite with non-collinear magnetic order.   The limiting case $\delta\to 0$ maps to a bipartite square lattice with collinear magnetic order and  $\delta\to \infty$ maps to a decoupled antiferromagnetic chains. For $\delta<1/2$  it has been established that the classical ground state is collinear (up-down-down state) and there is no degeneracy except a global spin rotation about the $z$-axis, and  for $\delta>1/2$ the classical ground state is non-collinear but coplanar. The collinear phases are trivial. In the present study, we focus on the canted coplanar/noncollinear regime $(\delta>1/2)$ as the ground state properties of volborthite, vesignieite, and edwardsite are believed to live in this regime.

\section{Classical Ground State}
 Now, we consider the classical ground state of the full Hamiltonian \eqref{model}.  In the classical limit, the spin operators  can be approximated as classical vectors,  written as
 $\bold{S}_{i}= S\bold{n}_i$, where $\bold{n}_i=\lb\sin\phi\cos\theta_i, \sin\phi\sin\theta_i,\cos\phi \rb$
 is a unit vector and $\theta_i$ labels the spin oriented angles on the spin triad and $\phi$ is the magnetic-field-induced canting angle. For $\delta>1/2$ the ground state is the canted coplanar/noncollinear spin configuration in Fig.~\ref{dis}. 
 The classical energy is given by
\begin{align}
e_0(\phi)&=2J(2+\delta)\big[\lb 1 -\cos\vartheta\rb\cos^2\phi+\cos\vartheta\big]\label{cla1}\\&\nonumber-4D_\perp\sin^2\phi\sin\vartheta(1-\cos \vartheta)-3h \cos\phi,
\end{align}
where $e_0(\phi)=E(\phi)/NS^2$, and $N$ is the number of sites per unit cell.  The magnetic field is rescaled in unit of $S$. The minimization of $e_0(\phi)$ yields the magnetic-field-induced canting angle $\cos\phi=h/h_s$ where \bea
h_s=\frac{\lb 1-\cos\vartheta\rb}{3}\big[4 J(2+\delta) + 8D_\perp\sin\vartheta\big].
\eea
 As can be seen from Eq.~\ref{cla1} the classical energy depends on the DM interaction as it contributes to the stability of the coplanar/noncollinear spin configuration. In collinear spin configurations, the DM interaction does not contribute to the classical energy, instead it provides TMDs, e.g. in kagom\'e ferromagnets  \cite{alex0, alex1, alex1a, jun, alex4, alex6a, alex6,alex44,alex2,mok,su, zhh,chen}.

\section{Topological Magnetic Excitations}
  \label{mag}
  The magnetic excitations of frustrated QKAF can be studied in different formalisms. In real materials the  intrinsic DM spin-orbit interaction and an external applied magnetic field are very likely to induce a magnetic long-range order in frustrated QKAF \cite{Yo,Yo1, Yo2,jeo,koz}. Therefore, we will employ  the Holstein-Primakoff (HP) transformation \cite{HP}  valid in the low temperature regime.   In the Supplemental Material (SM), we have shown that  a magnetic field (H $\perp$ 2D plane)  can induce a chiral noncoplanar spin texture with an emergent scalar spin chirality  \bea \mathcal H_{\phi}=J_{\phi}\sum_{i,j,k,\Delta}\chi_{ijk},\eea where $J_{\phi}=\cos\phi\propto h$. As we previously mentioned, the scalar spin chirality  breaks time-reversal ($\mathcal T$) symmetry macroscopically and can be spontaneously developed in CSL \cite{kal,wen,bau,bas, shou, shou1, shou2,shou3}. Its effects on the magnetic excitations should be the same in the ordered and disordered phases. In the magnetic-field-induced  noncoplanar regime  the DM interaction is not necessarily needed. Its main effect on QKAF is to induced a long-range magnetic order.  The system exhibits nontrivial topological effects through real space Berry curvature of the chiral noncoplanar spin texture. 

  \begin{figure}
\centering
\includegraphics[width=3in]{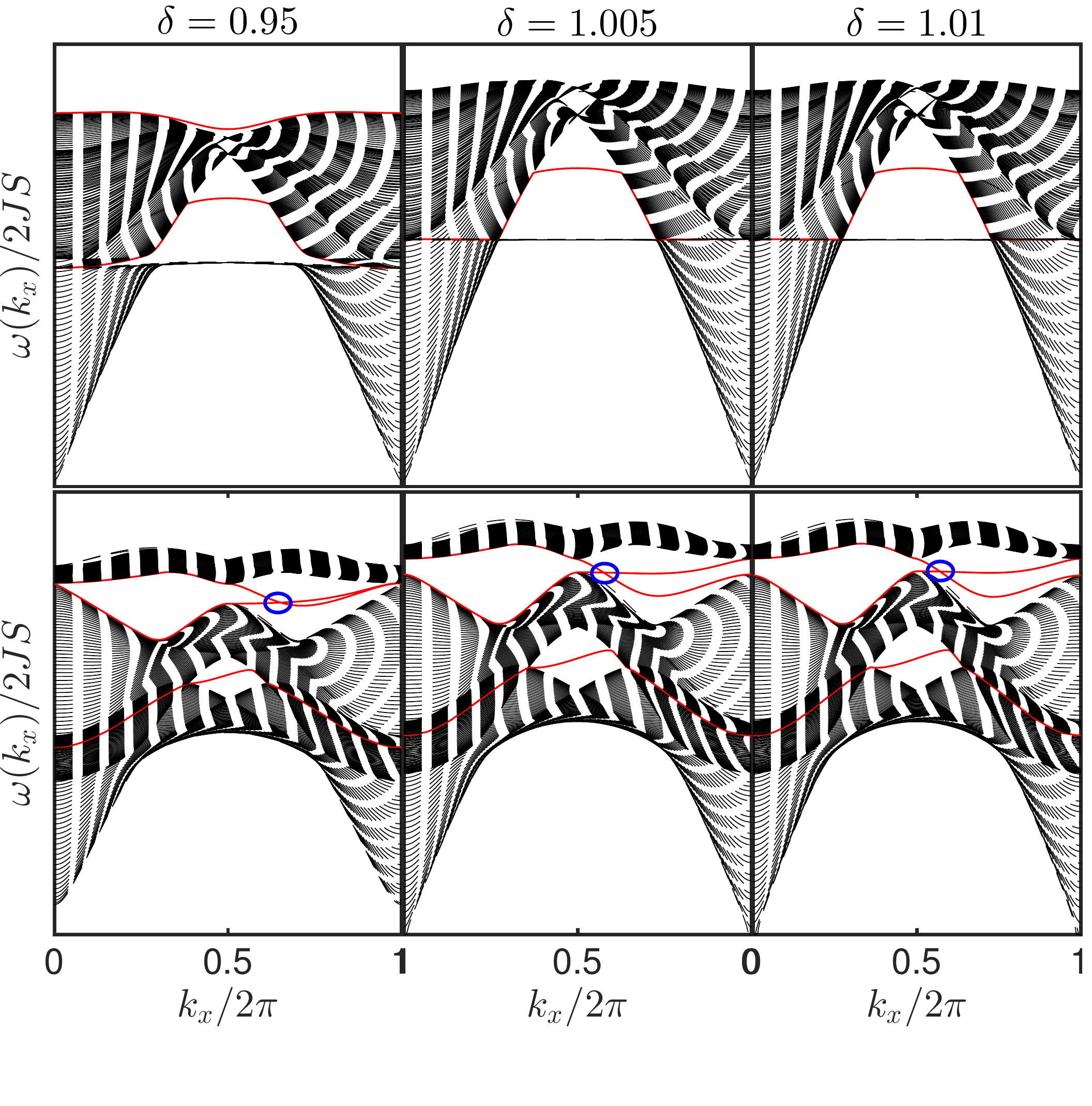}
\caption{Color online. Chiral  magnon edge modes (red lines) of distorted QKAF for a strip geometry with $D_\perp/J=0.2$. Top panel: $\mathcal{H}_\phi=0, ~h/h_s=0$. Bottom panel: $\mathcal{H}_\phi\neq 0, ~h/h_s=0.3$.}
\label{edge}
\end{figure} 

\begin{figure*}
\centering
\includegraphics[width=.8\linewidth]{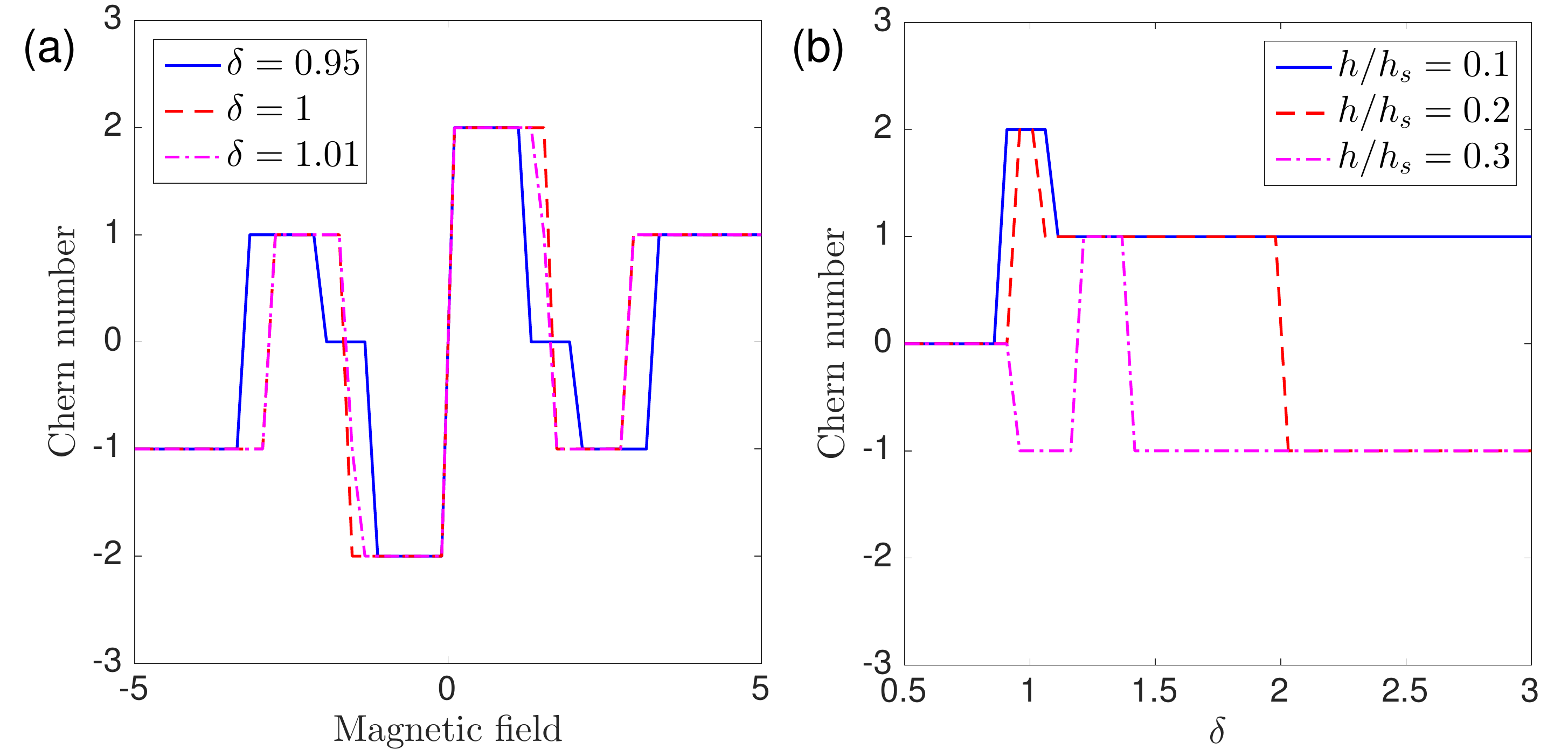}
\caption{Color online.  Variation of Chern number of the lowest magnon band as a function of (a) the magnetic-field and (b) the distortion for $D_\perp/J=0.2$.}
\label{chern}
\end{figure*}
  
The magnetic excitations at zero magnetic field  and zero DM interaction with $\delta\neq 1$  still exhibit a zero energy mode $\omega_0=0$ (not shown). Lattice distortion is unable to lift the zero energy mode. In addition, there are two non-degenerate dispersive modes $\omega_1\neq \omega_2$ (not shown)  induced by lattice distortion in contrast to undistorted kagom\'e antiferromagnets with degenerate modes $\omega_1= \omega_2$.   Although the exact parameter values of volborthite are not known, an out-of-plane DM interaction is intrinsic to the kagom\'e lattice. A moderate DM interaction ($D_\perp/J=0.2$) lifts the zero energy mode and stabilizes the coplanar/noncollinear spin configuration as shown in Fig.~\ref{band}. We find that the lifted zero mode is no longer a constant flat band but varies in the BZ (see SM), that is $\omega_0({\bf \Gamma})\neq \omega_0({\bf K})$, where ${\bf \Gamma}=(0,0)$ and ${\bf K}=(2\pi/3,~0)$ see Fig.~\ref{E0}. This is one of the differences  between distorted and undistorted kagom\'e antiferromagnets  at finite DM interaction.

 The symmetry group of the kagom\'e lattice is generated by lattice translation $\bold {T}$, $6$-fold rotation $\mathcal C_6$ ( $\pi/3$ rotation about the center of the hexagonal plaquettes), and mirror reflection symmetry $\mathcal M$ (see Fig.~\eqref{dis}). Several combinations of the space group symmetry and $\mathcal T$-symmetry can lead to identity in the coplanar order phase. For instance, $\mathcal M$-symmetry  is a good symmetry of the kagom\'e lattice, but it flips the in-plane spins and $\mathcal T$-symmetry flips the spins again and leaves the magnetic order invariant. Hence $\mathcal T\mathcal M$ is a symmetry of the coplanar/noncollinear magnetic order.   Since the lattice distortion  and  the out-of-plane DM interaction  preserve this combined symmetry the system should be topologically trivial as we will show below.  Therefore, we expect this model to be topologically nontrivial when either $\mathcal T$-symmetry or  $\mathcal M$-symmetry is broken.
    \begin{figure}
  \centering
\includegraphics[width=3in]{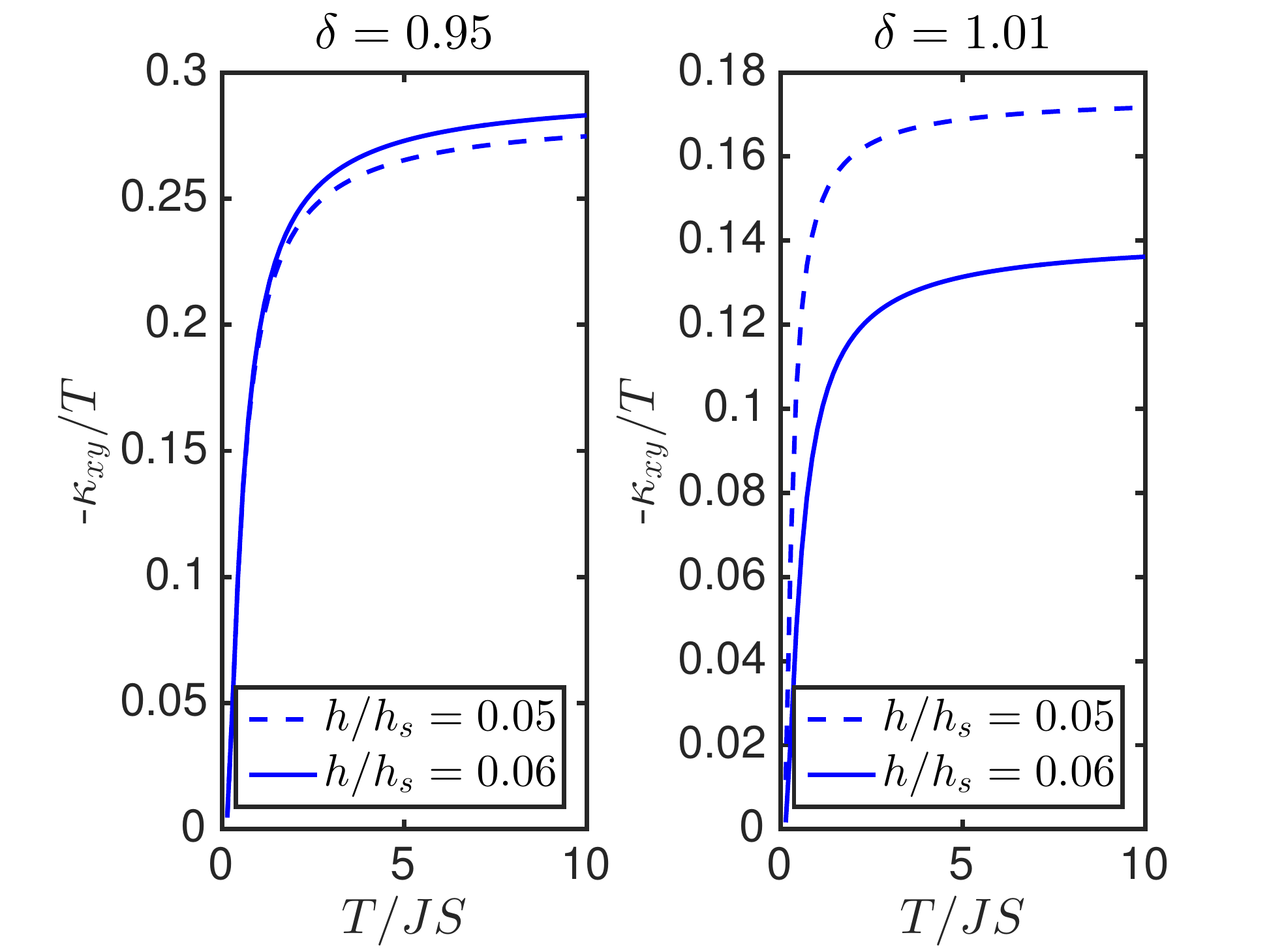}
\caption{Color online. Low temperature dependence of the topological thermal Hall conductivity   with $D_\perp/J=0.2$.  }
\label{hall}
\end{figure}
  \begin{figure}
  \centering
\includegraphics[width=3in]{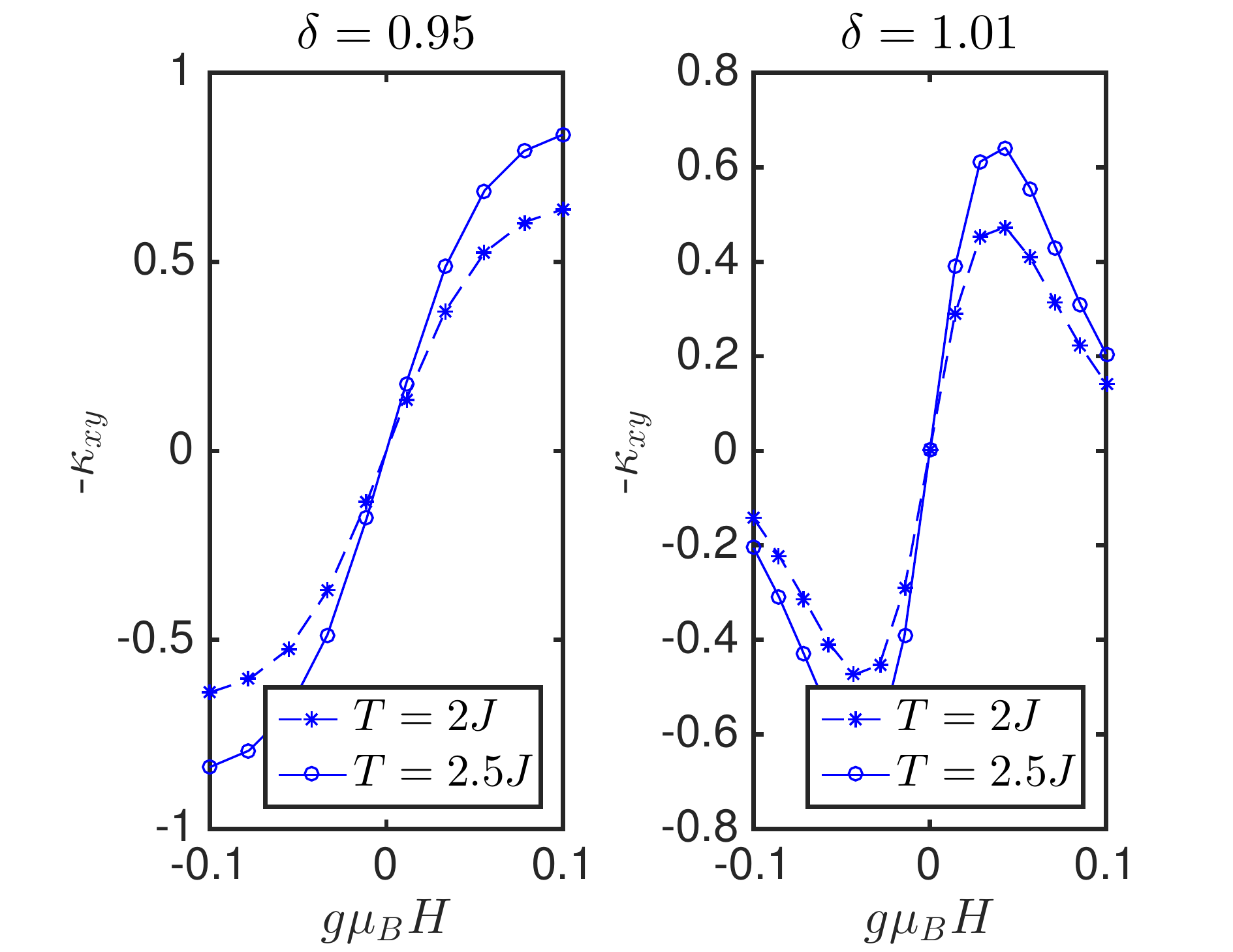}
\caption{Color online. Field dependence of the topological thermal Hall conductivity  at low temperatures   with $D_\perp/J=0.2$.  }
\label{hallm}
\end{figure}

The combined symmetry can be broken in two ways:

 $(1)$ If the kagom\'e lattice lacks  $\mathcal M$-symmetry, an in-plane DM component $D_{\parallel}$  can be allowed according to the Moriya rules \cite{dm1}. The in-plane DM component preserves $\mathcal T$-symmetry but breaks $\mathcal M$-symmetry, hence  $\mathcal T\mathcal M$ will be broken. This leads  to  weak out-of-plane ferromagnetism or noncoplanar spin canting with scalar spin chirality and weak ferromagnetic moment \cite{men1, mes1}. Thus, for kagom\'e antiferromagnetic materials with $D_{\perp}\ll D_{\parallel}$ we expect  the out-of-plane ferromagnetism to be dominant and hence the (in-plane) DM interaction should be  the primary source of topological spin excitations as we already know in collinear ferromagnets with out-of-plane DM interaction. However, most kagom\'e antiferromagnetic materials have a dominant intrinsic out-of-plane component $D_{\perp}\gg D_{\parallel}$ and the weak ferromagnetism induced by $D_{\parallel}$ can be negligible. In fact,  there was no signal of any DM interaction on the observed  $\kappa_{xy}$  in  kagom\'e volborthite \cite{wat}.  Therefore, there must be another source of topological spin excitations apart from the DM interactions.

  $(2)$ An alternative way to break $\mathcal T\mathcal M$ symmetry is by applying an external magnetic field perpendicular to the kagom\'e plane (H $\perp$ 2D plane). A finite out-of-plane magnetic field breaks $\mathcal T$-symmetry but preserves $\mathcal M$-symmetry.  It induces a noncoplanar spin texture with a nonzero scalar spin chirality (see SM).  The magnon dispersions are also gapped as depicted in Fig.~\ref{band1} similar to the case without magnetic field. However, the gap excitations with and without the magnetic field are different which can be shown by solving for the chiral edge modes.  The magnon edge modes  for a strip geometry with open boundary conditions along the $y$-direction  and infinite along $x$-direction are shown in Fig.~\ref{edge} at zero magnetic field (top panel). We see that the system does not possess topologically protected gapless edge modes and the Chern number vanishes (see SM). On the other hand, the edge modes at finite magnetic field shown  in bottom panel of Fig.~\ref{edge} are gapless --- an indication of a topological system.  We have also shown the variation of the Chern number of the lowest magnon band in Fig.~\ref{chern}.    The magnetic-field-induced noncoplanar chiral spin texture does not require a DM interaction in general as long-range order can be induced through other interactions such as a second-nearest neighbour. Besides, the scalar spin chirality $\chi_{ijk}$ should have the same effects on magnetic excitations in the CSL.

\section{Topological Thermal Hall Effect}
The mechanism that gives rise to a nonzero thermal Hall effect in insulating frustrated magnets  has remained an open question. In electronic (metallic) systems there is a concept of the topological Hall effect in which  an unconventional anomalous Hall conductivity is induced by the scalar spin chirality resulting from the configuration of the chiral spin textures  rather than the spin-orbit interaction as reported in several electronic materials \cite{ele0,ele1,ele2,ele}. In this section, we will show that this mechanism can be observed also in charge-neutral magnetic excitations on the frustrated QKAF. In this case the appropriate transport measurement is  the thermal Hall effect in   which a temperature gradient $-\boldsymbol\nabla T$  induces a heat current $\boldsymbol {\mathcal J}^Q$. From linear response theory, one obtains $\mathcal J_{\alpha}^Q=-\sum_{\beta}\kappa_{\alpha\beta}\nabla_{\beta} T$, where $\kappa_{\alpha\beta}$ is the thermal conductivity  and the transverse component $\kappa_{xy}$ is associated with the thermal Hall conductivity given explicitly in Ref.~\cite{shin1}.
 
Figure \ref{hall} shows the low-temperature dependence of $\kappa_{xy}$ (in units of $k_B/\hbar$)  for two values of the lattice distortion and the magnetic field.  The  topological Hall conductivity  captures  a negative value in both regimes $\delta<1$ and  $\delta>1$ and the thermal Hall conductivity is suppressed in the latter.  As we previously noted the thermal Hall conductivity does not originate from the DM spin-orbit interaction but from the chiral spin texture of the noncoplanar spins induced by the magnetic field. The DM spin-orbit interaction is only necessary to stabilize the coplanar order but a second-nearest neighbour antiferromagnetic interaction or an easy-plane anisotropy  also stabilizes the coplanar/noncollinear N\'eel order.   Therefore, kagom\'e antiferromagnetic materials with  a weak (negligible) DM interaction can possess a thermal Hall conductivity from the real space Berry curvature of the chiral spin texture and this should be  possible in the CSL phase even at zero magnetic field. This is an analog of topological Hall effect   in electronic systems \cite{ele0,ele1,ele2,ele}.
 On the other hand, Fig.  \ref{hallm} shows the magnetic field  dependence of $\kappa_{xy}$ which shows a symmetric sign change as the magnetic field is reversed as can be understood from the rotation of the spins and the corresponding sign change in the spin chirality.  We note that the role of magnetic field in the QKAF is different from the triplon bands in a dimerized quantum magnet  \cite{jud}, where no scalar spin chirality was induced $\chi_{ijk}=0$, therefore topological magnetic excitations and thermal Hall effect stem from the DM interaction \cite{jud}.


\section{Conclusion}
 In summary, we have studied topological magnetic excitations and thermal Hall effect induced by the real space Berry curvature of the chiral spin texture on the distorted kagom\'e antiferromagnets  applicable to vesignieite, edwardsite, and volborthite.  The lack of inversion symmetry on the kagom\'e lattice allows a Dzyaloshinskii-Moriya (DM)  interaction. However, in contrast to ferromagnets the DM interaction is only necessary to stabilize  the coplanar/noncollinear magnetic spin structure. We showed that lattice distortion and DM interaction introduce gap magnetic excitations, but the system remains topologically trivial with neither protected chiral edge modes nor finite thermal Hall conductivity. By turning on an out-of-plane external magnetic field, we showed that a finite scalar spin chirality is induced from the noncoplanar chiral spin texture. This gives rise to a real space Berry curvature which measures the solid angle subtended by three noncoplanar spins independent of the DM interaction. We note that the real space Berry curvature of the chiral spin texture should persist in the chiral spin liquid (CSL) phase due to the presence of the spontaneous scalar spin chirality.

The results of this Letter suggest that the experimental result of thermal Hall response in volborthite can be best described using a distorted Heisenberg  model as opposed to a frustrated alternating spin chain.  It is noted that the DM interaction  can be removed by a gauge transformation  for frustrated one-dimensional (1D) spin chain \cite{zyu} and an ideal 1D system should not have a chiral noncoplanar magnetic order and scalar spin chirality should be absent.  An alternative approach is the Schwinger boson formalism, however the  scalar spin chirality does not appear explicitly  in this formalism. Instead  the DM interaction generates a magnetic flux   leading to topological spin excitations even in the absence of an applied magnetic field \cite{mes,alex44}. This is reminiscent of collinear ferromagnets and thus  sharply contrast with the present results.  

 These results also call for further investigation of thermal Hall effect in the kagom\'e volborthite and other  distorted quantum kagom\'e antiferromagnets such as vesignieite  \cite{oka,oka1,oka2,okaa,okaa1, men0, men2} and  edwardsite   \cite{oka3}. Although we captured a negative thermal Hall conductivity ($\kappa_{xy}$) as seen in experiment \cite{wat}, it would be interesting to determine the parameter values of volborthite  and also measure the magnetic field dependence of $\kappa_{xy}$ and possibly the topological magnetic excitations at various magnetic field and temperature ranges.  Furthermore, experiment should also try to measure the spontaneous or magnetic-field-induced scalar spin chirality that gives rise to topological magnetic excitations and thermal Hall response.  We believe that the results of this Letter have shed some light on recent observation of thermal Hall conductivity in volborthite \cite{wat}.

\acknowledgements
The author would like to thank  M. Yamashita,  for elaborating on the experimental result of thermal Hall response in  kagom\'e volborthite. Research at Perimeter Institute is supported by the Government of Canada through Industry Canada and by the Province of Ontario through the Ministry of Research
and Innovation.

\end{document}


\date{\today}
\title{Topological Magnetic Excitations   on the Distorted Kagom\'e Antiferromagnets\\Supplemental Material}
\author{S. A. Owerre}
\affiliation{Perimeter Institute for Theoretical Physics, 31 Caroline St. N., Waterloo, Ontario N2L 2Y5, Canada.}
\affiliation{African Institute for Mathematical Sciences, 6 Melrose Road, Muizenberg, Cape Town 7945, South Africa.}

\maketitle

\section{Model Hamiltonian}
\label{appena}
We consider the  Hamiltonian for kagom\'e antiferromagnets with DM interaction and Zeeman magnetic field given by
 \begin{align}
\mathcal H&= \sum_{\la i,j\ra}\big [ J_{ij}{\bf S}_{i}\cdot{\bf S}_{j}+ { \bf D}_{ij}\cdot{ \bf S}_{i}\times{\bf S}_{j}\big]-{\bf H}\cdot\sum_{i} {\bf S}_{i},
\label{h}
\end{align}
where ${\bf H}=h{\bf \hat z}$,  $h=g\mu_B H$ and $\bold{D}_{ij}=-D_\perp{\bf \hat z}$.  This particular case of out-of-plane DM interaction is  very common in most distorted kagom\'e antiferromagnetic materials.  It is also  interesting because topological magnetic excitations are not induced by the DM interaction as we will show. The presence of out-of-plane DM interactions breaks the complete SU(2) rotation symmetry down to U(1) symmetry about the $z$-axis.   The classical energy is given by
\begin{align}
e_0(\phi)&=2J(2+\delta)\big[\lb 1 -\cos\vartheta\rb\cos^2\phi+\cos\vartheta\big]\label{cla1}\\&\nonumber-4D_\perp\sin^2\phi\sin\vartheta(1-\cos \vartheta)-3h \cos\phi,
\end{align}
where $e_0(\phi)=E(\phi)/NS^2$, $N$ is the number of sites per unit cell, and $\vartheta= \arccos(-1/2\delta)$ which is not exactly $120^\circ$ for $\delta\neq 1$. The magnetic field is rescaled in unit of $S$. The minimization of $e_0(\phi)$ yields the field-induced canting angle $\cos\phi=h/h_s$ where \bea
h_s=\frac{\lb 1-\cos\vartheta\rb}{3}\big[4 J(2+\delta) + 8D_\perp\sin\vartheta\big].
\eea
The  excitations above the classical ground state are obtained as follows. The procedure involves  performing a rotation  about the $z$-axis on the triad by the spin oriented  angles $\vartheta$ in order to achieve the  coplanar configuration. As the out-of-plane magnetic field is turned on, we have to align the spins along the new quantization axis by performing a rotation about the $y$-axis by the field canting angle $\phi$.  The total rotation matrix takes the form \bea \bold{S}_i=\mathcal{R}_z(\theta_i)\cdot\mathcal{R}_y(\phi)\cdot\bold S_i^\prime,\eea
where
\begin{align}
\mathcal{R}_z(\theta_i)\cdot\mathcal{R}_y(\phi)
=\begin{pmatrix}
\cos\theta_i\cos\phi & -\sin\theta_i & \cos\theta_i\sin\phi\\
\sin\theta_i\cos\phi & \cos\theta_i &\sin\theta_i\sin\phi\\
-\sin\phi & 0 &\cos\phi
\end{pmatrix},
\label{rot}
\end{align}
and $\theta_i=\vartheta_{A,B,C}$.
In the following, we drop the prime in the rotated coordinate. At low temperatures accessible experimentally, the noninteracting magnon model sufficiently describes the system. The corresponding Hamiltonian that contribute to noninteracting magnon model is given by
  \begin{align}
 &\mathcal H_{J}=   \sum_{\la i, j\ra}J_{ij}\big[\cos\theta_{ij} \bold S_{i}\cdot \bold S_{j}+ \sin\theta_{ij}\cos\phi \hat{\bf z}\cdot \lb \bold S_{i}\times\bold S_{j}\rb  \\&\nonumber +2\sin^2\lb\frac{\theta_{ij}}{2}\rb(\sin^2\phi S_{i}^xS_{j}^x +\cos^2\phi S_{i}^zS_{j}^z)\big], \\
&\mathcal   H_{DM}=  D_\perp\sum_{\la i, j\ra}\big[ \sin\theta_{ij}(\cos^2\phi S_{i}^xS_{j}^x + S_{i}^yS_{j}^y\\&\nonumber+\sin^2\phi S_{i}^zS_{j}^z)-\cos\theta_{ij}\cos\phi \hat{\bf z}\cdot\lb \bold S_{i}\times\bold S_{j}\rb\big], \\
 &\mathcal H_h=-h\cos\phi\sum_{i} S_{i}^z,
     \end{align}
 where $\theta_{ij}=\theta_i-\theta_j$.  At zero magnetic field, i.e., $\phi=\pi/2$, the chiral interaction is absent in the magnon model despite the the presence of DM interaction. Hence, the system is topologically trivial at zero field. The magnetic-field-induced scalar spin chirality  \bea \mathcal H_{\chi}= \sum_{i,j,k=\Delta} \chi_{ijk},\eea
  originates from noncoplanar spin texture formed by the spin triad ${\bold S}_i,~{\bold S}_j$, ${\bold S}_k$, where $\chi_{ijk}=  {\bold S}_i\cdot\lb \bold S_{j}\times\bold S_{k}\rb$. It is well-known that $\la \chi_{ijk}\ra$ can be nonzero even in the absence of magnetic ordering $\la {\bf S}_j\ra=0$, e.g. in chiral spin liquid phase.  This is a very crucial difference between collinear ferromagnets and antiferromagnets on the kagom\'e lattice.  To obtain the magnon dispersions,  we proceed as usual by introducing the  Holstein Primakoff spin bosonic operators  
\begin{align} 
& S_{i}^{z}= S-a_{i}^\dagger a_{i},
\\& S_{i}^{ y}=  i\sqrt{\frac{S}{2}}(a_{i}^\dagger -a_{i}),
\\& S_{i}^{ x}=  \sqrt{\frac{S}{2}}(a_{i}^\dagger +a_{i}),
\end{align}
 where $a_{i}^\dagger(a_{i})$ are the bosonic creation (annihilation) operators. 
 The noninteracting magnon tight binding Hamiltonian is given by
\begin{align}
\mathcal H&= S\sum_{\la i,j\ra}\big[G_{ij}^z(a_i^\dg a_i +a_j^\dg a_j) +G_{ij}^d (e^{-i\Phi_{ij}}a_i^\dg a_j + h.c.)\\&\nonumber +G_{ij}^{o}(a_i^\dg a_j^\dg + h.c.)\big]+h_{\phi}\sum_{i}a_i^\dg a_i,
\end{align}
where
\begin{align}
&G_{ij}^z=-J_{ij}[\cos\theta_{ij}+2\cos^2\phi\sin^2(\theta_{ij}/2)]\\&\nonumber- \mathcal D_\perp\sin^2\phi
\sin{\theta_{ij}},\\
&G_{ij}^d=\sqrt{(G^R_{ij})^2+(G^M_{ij})^2},\\
&G^R_{ij}= J_{ij}\big[\cos\theta_{ij} +\frac{2\sin^2\phi\sin^2\lb\theta_{ij}/2\rb}{2}\big]\\&\nonumber+ D_\perp\sin\theta_{ij}\lb 1-\frac{\sin^2\phi}{2}\rb,\\
&G^M_{ij}= \cos\phi\lb  J_{ij}\sin\theta_{ij}- D_\perp\cos\theta_{ij}\rb,\\
&G_{ij}^{o}=\frac{\sin^2\phi}{2}\lb 2 J_{ij}\sin^2(\theta_{ij}/2) - D_\perp\sin\theta_{ij}\rb,
\end{align}
and $h_\phi=h\cos\phi$. The fictitious magnetic flux or solid angle subtended by three noncoplanar spins  is given by $\tan\Phi_{ij}=G^M_{ij}/G^R_{ij}$. We clearly see that $\Phi_{ij}$ is nonzero in the absence of DM interaction.  It should be noted that the $\bold{ Q=0}$ magnetic structure that can be stabilized by other means as mentioned in the text. In momentum space we obtain  \begin{align}
&\mathcal H= S\sum_{\bo,\alpha,\beta}\lb 2\mathcal{M}_{\alpha\beta}^z\delta_{\alpha\beta} +2\mathcal{M}_{\alpha\beta}^d\rb a_{\bo \alpha}^\dagger a_{\bo \beta}\label{main}\\&\nonumber +\mathcal{M}_{\alpha\beta}^{o} \lb a_{\bo \alpha}^\dagger a_{-\bo \beta}^\dagger +a_{\bo \alpha} a_{-\bo \beta}\rb,
\end{align}
where $\alpha,\beta=A,~B,~C$ and the coefficients are given by

\bea
\boldsymbol{\mathcal{M}}^z=\text{diag}\lb \zeta_{AA},\zeta_{BB},\zeta_{CC}\rb,
\eea
with $\zeta_{AA}=G_{AB}^z+G_{CA}^z +h_\phi/2$, $\zeta_{BB}=\zeta_{CC}=G_{AB}^z+G_{BC}^z +h_\phi/2$. 
\begin{align}
&\boldsymbol{\mathcal{M}}^d=\begin{pmatrix}
  0& \gamma_{AB}^de^{-i\Phi_{AB}}&\gamma_{CA}^de^{i\Phi_{CA}} \\
\gamma_{AB}^{*d} e^{i\Phi_{AB}}& 0&\gamma_{BC}^{d}e^{-i\Phi_{BC}}\\
\gamma_{CA}^{*d} e^{-i\Phi_{CA}}& \gamma_{BC}^{*d} e^{i\Phi_{BC}} & 0 \\  
 \end{pmatrix},\\
 &\boldsymbol{\mathcal{M}}^o=\begin{pmatrix}
  0& \gamma_{AB}^o &\gamma_{CA}^o \\
\gamma_{AB}^{*o} & 0&\gamma_{BC}^o\\
\gamma_{CA}^{*o} & \gamma_{BC}^{*o} & 0 \\  
 \end{pmatrix},
\end{align}
where $\gamma_{AB}^d= G_{AB}^d \cos k_1, ~\gamma_{BC}^d= G_{BC}^d \cos k_2, ~\gamma_{CA}^d=G_{CA}^d \cos k_3;~\gamma_{AB}^o=G_{AB}^o\cos k_1,~ \gamma_{BC}^o=G_{BC}^o\cos k_2,  \gamma_{CA}^o=G_{CA}^o \cos k_3$.  $k_i=\bold k_i\cdot \bold e_i$, and $\bold e_1=(-1/2,~-\sqrt 3/2),~\bold e_2=(1,0),~\bold e_3=(-1/2,~\sqrt 3/2)$.
 The Hamiltonian can be written as 
\bea \mathcal H=S\sum_{\bo} \Psi^\dg_\bo \boldsymbol{ \mathcal{H}}(\bo)\Psi_\bo-\text{const.},\eea
where  $\Psi^\dg_\bo= (b_{\bo A}^{\dg},\thinspace b_{\bo B}^{\dg},\thinspace b_{\bo C}^{\dg}, \thinspace b_{-\bo A},\thinspace b_{-\bo B},\thinspace b_{-\bo C} )$, and \begin{align}
&\boldsymbol{ \mathcal{H}}(\bo)= \begin{pmatrix}
  {\boldsymbol{\mathcal{M}}^{z}}+\boldsymbol{\mathcal{M}}^d& \boldsymbol{\mathcal{M}}^o\\
\boldsymbol{\mathcal{M}}^o &\  {\boldsymbol{\mathcal{M}}^{z}}+\boldsymbol{\mathcal{M}}^d\\  
 \end{pmatrix}.
\end{align}
In the undistorted limit $\delta\to 1$, $\zeta_{AA}=\zeta_{BC}=\zeta_{CA}$. The resulting magnon bands at zero field $\mathcal H_{\chi}=0$ has a flat constant  mode.  For the distorted case $\delta\neq 1$ the situation is different. We have $\zeta_{AA}>\zeta_{BC},~\zeta_{CA}$ for $\delta<1$ and $\zeta_{AA}<\zeta_{BC},~\zeta_{CA}$ for $\delta>1$. The magnon bands at zero field $\mathcal H_{\chi}=0$ no longer have  a flat constant  mode in the entire Brillouin zone see main text.

\section{Matrix diagonalization}
The Hamiltonian is diagonalized by the generalized Bogoliubov transformation 
$\Psi_\bo= \mathcal{P}_\bo Q_\bo$, 
where $\mathcal{P}_\bo$ is a $2N\times 2N$ paraunitary matrix and  $Q^\dg_\bo= (\mathcal{Q}_\bo^\dg,\thinspace \mathcal{Q}_{-\bo})$ with $ \mathcal{Q}_\bo^\dg=(\beta_{\bo A}^{\dg}\thinspace \beta_{\bo B}^{\dg}\thinspace \beta_{\bo C}^{\dg})$ being the quasiparticle operators. The matrix $\mathcal{P}_\bo$ satisfies the relations,
\begin{align}
&\mathcal{P}_\bo^\dg \boldsymbol{\mathcal{H}}(\bo) \mathcal{P}_\bo=\mathcal{E}_\bo\label{eqn1}\\ &\mathcal{P}_\bo^\dg \boldsymbol{\tau}_3 \mathcal{P}_\bo= \boldsymbol{\tau}_3,
\label{eqna}
\end{align}
where $\mathcal{E}_\bo= \text{diag}(\omega_{\bo\alpha},~\omega_{-\bo\alpha})$, 
$ \boldsymbol{\tau}_3=
\text{diag}(
 \mathbf{I}_{N\times N}, -\mathbf{I}_{N\times N} )$, and $\omega_{\bo\alpha}$ are the  energy eigenvalues.
From Eq.~\ref{eqna} we get $\mathcal{P}_\bo^\dg= \boldsymbol{\tau}_3 \mathcal{P}_\bo^{-1} \boldsymbol{\tau}_3$, and Eq.~\ref{eqn1} is equivalent to saying that we need to diagonalize the Hamiltonian $\boldsymbol{\mathcal{H}}^\prime(\bo)= \boldsymbol{\tau}_3\boldsymbol{\mathcal{H}}(\bo),$ whose eigenvalues are given by $ \boldsymbol{\tau}_3\mathcal E_\bo$ and the columns of $\mathcal P_\bo$ are the corresponding eigenvectors. The eigenvalues of this Hamiltonian cannot be obtained analytically except at zero field.  The paraunitary operator  $\mathcal P_\bo$ defines a Berry curvature given by
 \begin{align}
 \Omega_{ij;\alpha}(\bo)=-2\text{Im}[ \boldsymbol{\tau}_3\mathcal (\partial_{k_i}\mathcal P_{\bo\alpha}^\dg) \boldsymbol{\tau}_3(\partial_{k_j}\mathcal P_{\bo\alpha})]_{\alpha\alpha},
 \label{bc1}
 \end{align}
with $i,j=\lbrace x,y\rbrace$ and $\mathcal P_{\bo\alpha}$ are the columns of $\mathcal P_{\bo}$. In this form, the Berry curvature simply extracts the diagonal components which are the most important.  From Eq.~\ref{eqn1} the Berry curvature can be written alternatively as

\begin{align}
\Omega_{ij;\alpha}(\bold k)=-2\sum_{\alpha^\prime\neq \alpha}\frac{\text{Im}[ \braket{\mathcal{P}_{\bo\alpha}|v_i|\mathcal{P}_{\bo\alpha^\prime}}\braket{\mathcal{P}_{\bo\alpha^\prime}|v_j|\mathcal{P}_{\bo\alpha}}]}{\lb\omega_{\bo\alpha}-\omega_{\bo\alpha^\prime}\rb^2},
\label{chern2}
\end{align}
where $\bold v=\partial\boldsymbol{\mathcal{H}}^\prime(\bo)/\partial \bold k$ defines the velocity operators. The Berry curvature is related to the solid angle $\Omega(\bold k)\propto \Phi$ and the Chern number is defined as,
 \begin{equation}
\mathcal{C}_\alpha= \frac{1}{2\pi}\int_{{BZ}} dk_xdk_y~ \Omega_{xy;\alpha}(\bold k).
\label{chenn}
\end{equation}  
It can be shown that the Chern numbers are related to the scalar-chirality-induced fictitious flux $\Phi$ as $\mathcal{C}_1=0,~\mathcal{C}_2=-\text{sgn}(\sin(\Phi)),~\mathcal{C}_3=\text{sgn}(\sin(\Phi))$, where \bea\sin(\Phi)=\frac{1}{2}\chi_{ijk}.\eea